\documentclass[
    ,final            
  ]
  {aipproc}

\layoutstyle{8x11single}

\begin{document}

\title{Short-Range Nucleon-Nucleon Correlations}


\author{Douglas W. Higinbotham}{
  address={Thomas Jefferson National Accelerator Facility, Newport News, VA 23601, USA}
}

\begin{abstract}
Valence-shell nucleon knock-out experiments, such as $^{12}$C(e,e'p)$^{11}$B, measure
less strength then is predicted by independent particle shell model calculations.
The theoretical solution to this problem is to include the correlations 
between the nucleons in the nucleus in the calculations.  Motivated by these results, many 
electron scattering experiments have tried to isolate the signal from  
these correlations in order to gain new insight into the short-range part 
of the nucleon-nucleon potential. 
Unfortunately, many competing mechanisms can cause the same observable final-state as
an initial-state correlation, making truly isolating the signal extremely challenging.
This paper reviews the recent experimental evidence for short-range correlations,
as well as explores the possibility that such correlations are responsible
for the EMC effect in the $0.3 < x_B < 0.7$ deep inelastic scattering ratios.
\end{abstract}

\maketitle


\section{Introduction}

The independent particle shell model enjoyed many successes, but
failed to predict the strength of valance-shell nucleon knock-out 
experiments~\cite{Lapikas:1003zz,Kelly:1996hd}.  
The theoretical solution to this problem was simply to include the correlations that
must exist between the nucleons in the nucleus.
When compared to an early independent particle shell model calculations, these correlations
move strength away from the bound-states and into a continuum of energies and momentums.
This effect is illustrated in the work of Benhar {\it et al.} 
where it is shown how the $^{12}$C momentum distribution from an independent
particle model calculation is reduced at low momentum and enhanced at momentums above $\sim$250~MeV/c
by the addition of nucleon-nucleon correlations~\cite{Benhar:2006wy}. 
An \textit{ab-initio} calculation of the depletion of the nuclear Fermi sea by correlations 
can be found in Ref.~\cite{Rios:2009gb}.

\begin{figure}[ht]
  \includegraphics[width=\textwidth]{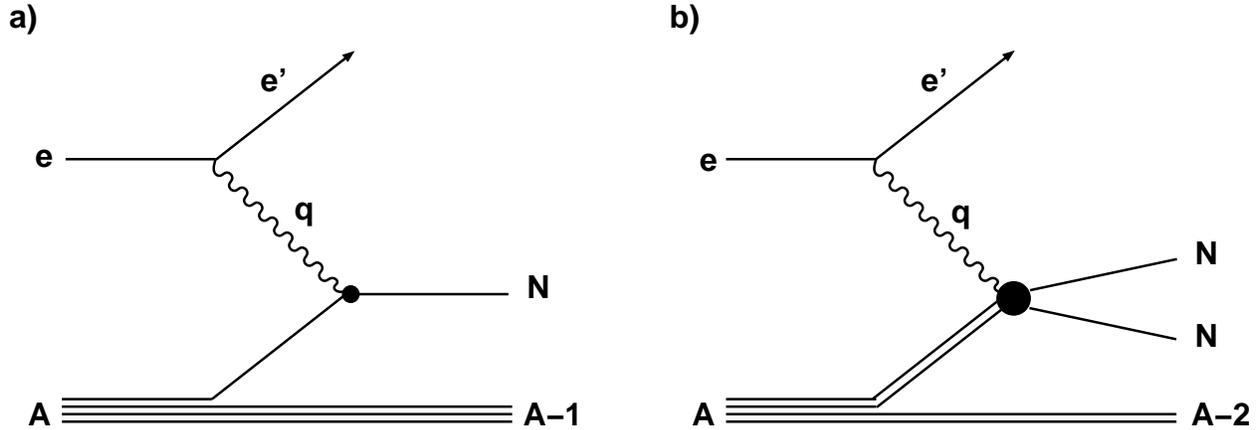}
  \caption{The simple goal of short-range nucleon-nucleon correlation studies is to cleanly isolate
diagram b) from a).  Unfortunately, there are many other diagrams, including those with final-state
interactions, that can produce the same final state as the diagram scientists would like to isolate.
If one could find kinematics that were dominated by diagram b) it would finally allow
electron scattering to provide new insights into the short-range part of the nucleon-nucleon potential.}
  \label{diagrams}
\end{figure}

If one could cleanly isolate the high momentum part of the momentum distribution,
it would allow electron scattering to provide
new information about the short-range part of the nucleon-nucleon potential.
This is shown diagrammatically in Fig.~\ref{diagrams} where a) shows simple single
nucleon knock-out and b) shows the electron knocking-out a correlated pair of nucleons..
Unfortunately, 
it has proven experimentally challenging to cleanly isolate the effects of short-range correlations
from other effects such as meson-exchange currents, isobar-configurations, and final-state
interactions.  Thus, while the 1981 D(e,e'p)n momentum distribution extraction up to $\approx$250~MeV/c
has stood the test of time~\cite{Bussiere:1981mv}, that same paper's prediction that one could probe the short-range 
part of the nucleon-nucleon potential by simply extending the range of kinematics has proven
extremely challenging.  

\section{Electron Scattering Kinematics}

The story of the search for nucleon-nucleon correlations is best told with the Brojken $x_B$ variable:
\begin{equation}
x_B = \frac{Q^2}{2m\omega}
\end{equation}
where $m$ is the mass of the proton, $\omega$ is the energy transfer given 
by the difference between the beam energy and the scattered electron energy,
Q$^2$ is the four-momentum transfer given by {\bf q}$^2$ - $\omega^2$ where
{\bf q} is the three-momentum transfer vector given by the difference between
the incident and scattered electron momentum vectors.
While this kinematic variable is typically used for deep inelastic physics,
it turns out to be a convent variable for counting how many nucleons must be involved in a
knock-out reaction.  For example, while elastic scattering on a proton has a $x_B$ = 1,
quasi-elastic scattering of a nucleon from a nucleus can kinematically 
extended into the $x_B > 1$ region due to the presence of the second nucleon.
For the A(e,e') reaction, the $x_B$ and Q$^2$ variables can be determined by 
knowing the properties of the incident beam
and measuring the energy and momentum of the scattered electron.

For A(e,e'p) reactions, one can determine not only the energy and moment transferred, but
also the energy and momentum of the knocked-out nucleon.  The difference between the transferred
and detected energy and momentum is referred to as the missing energy, $E_{miss}$ 
and missing momentum, $p_{miss}$, respectively.
From the theoretical works on how short-range nucleon-nucleon correlations
effects the momentum distribution of nucleons in the nucleus~\cite{CiofidegliAtti:1995qe}, 
it is clear one must probe beyond the simple particle in an average potential
motion of the nucleon in the nucleus of approximately 250~MeV/c in order to observe
the effects of correlations.

With the construction of the Jefferson Lab Continuous Electron Beam Facility (CEBAF)~\cite{Leemann:2001dg},
it was possible to do high-luminosity knock-out reactions in {\it ideal} 
quasi-elastic kinematics into the $p_{miss} > 250$~MeV/c region.
In the early Jefferson Lab knock-out reaction proposals, such as E89-044 $^3$He(e,e'p)pn and $^3$He(e,e'p)d, these
kinematics were argued as the key to cleanly observe the effects of short-range correlations.  
And while final results of the experiments were clearly effected by the presence of correlations, 
the magnitude of the cross sections in the high missing momentum region was dominated by final-state interaction 
effects~\cite{Benmokhtar:2004fs,Rvachev:2004yr}.  Equally striking was the D(e,e'p)n data 
from CLAS taken at Q$^2$ > 5 [GeV/c]$^2$ in $x_B < 1$ kinematics~\cite{Egiyan:2007qj}.
Here it was shown that meson-exchange currents, final-state interaction, and delta-isobar configurations
mask cleanly probing nucleon-nucleons even at extremely high Q$^2$ in $x_B < 1$ kinematics.

\section{Nuclear Scaling}

With both the $x_B < 1$ and $x_B = 1$ kinematics practically ruled out for ever being able to cleanly probe 
short-range correlations; there is only one region left to explore: $x_B > 1$.  This is a special region, since
it is kinematically forbidden for a free nucleon, and thus seems to be a natural place to observe effects of
multi-nucleon interactions.  These kinematics were probed with limited statistics at SLAC~\cite{Day:1990mf} and
the plateaus in the per nucleon ratios, r(A/d), were claimed at to be 
evidence for short-range correlations~\cite{Frankfurt:1993sp}. 

\begin{figure}[htb]
  \includegraphics[width=\textwidth]{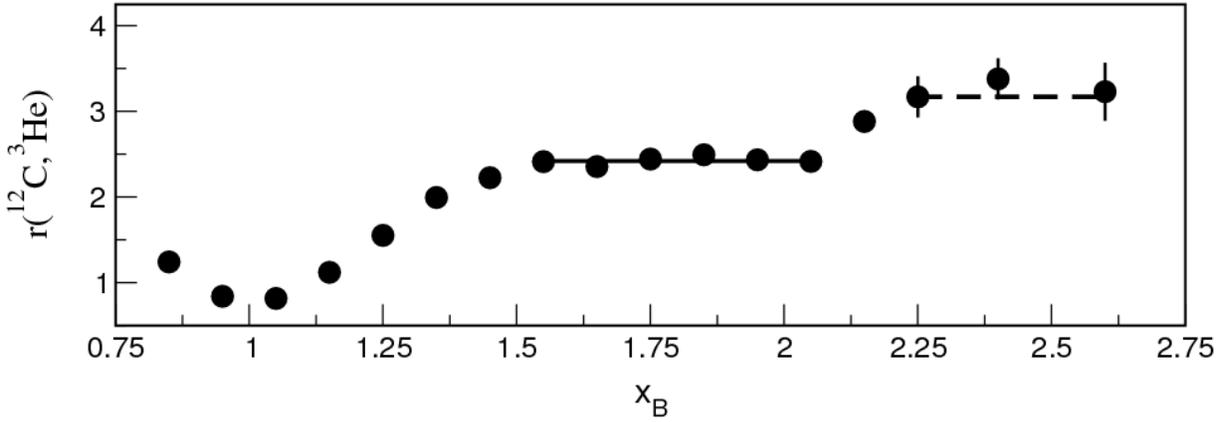}
  \caption{Shown is the per nucleon ratio of $^{12}$C(e,e') to $^3$He(e,e') 
           as a function of $x_B$ from CLAS~\cite{Egiyan:2003vg,Egiyan:2005hs}.  The data
           show two scaling plateaus.  The first, in the $x_B > 1$ region, is claimed to be 
           due to two-nucleon correlations; while the second, in the $x_B > 2$ region,
           is claimed to be due to three-nucleon correlations.}
  \label{CLAS-scaling}
\end{figure}

In 2003, CLAS published high statics data in the same kinematic region.  The results clearly showed that
the plateaus could only be seen for Q$^2$ > 1 [GeV/c]$^2$ and $x_B > 1$ kinematics~\cite{Egiyan:2003vg}
as predicted by Frankfurt and Strikman~\cite{Frankfurt:1981mk}.  
But plateaus alone are not evidence for correlations, just evidence that the functional form of the cross
section is the same for the two nuclei; so data was taken the $x_B > 2$ region.  By logic, if $1 < x_B < 2$
is a region of two-nucleon correlations, then the $x_B > 2$ region should be dominated by three-nucleon
correlations.  The CLAS Q$^2$ > 1 and $x_B > 2$ experiment reported observing a second 
scaling plateau as shown in Fig.~\ref{CLAS-scaling}~\cite{Egiyan:2005hs}.
Preliminary results of Hall C high precision data have shown roughly the same 
magnitude for these plateaus as CLAS and
shown that there is no Q$^2$ dependence in the $2 < Q^2 < 4$~[GeV/c]$^2$ range~\cite{Fomin:2008fq,Fomin:2008iq}. 

\section{Exclusive Reactions}

While the inclusive results alone are rather convincing, one would like to observe that the final-state
is truly dominated by a two-nucleon knock-out process.  This can be done by measuring two high-momentum
nucleons in the final-state and showing that the rest of the nuclear system is left nearly at rest. 
Two independent experiments have
been carried out that have done this, one at Brookhaven with a proton beam and another at Jefferson Lab with
an electron beam.  

The Brookhaven experiment made use of a proton beam on a carbon target and detected two forward
going protons.  Summing the momentum of the two protons and subtracting the incident proton's momentum,
allowed them to determined the missing momentum.  They then looked for recoiling neutrons with a large
acceptance detector.  For missing momentum below 250~MeV/c,
an isotropic distribution of neutrons were observed; but above 250~MeV/c the signal was completely
dominated by neutrons traveling in the direction of the 
missing momentum vector~\cite{tang:2002ww,Watson:2008zz}.  
Analysis of this result showed the reaction was dominated by 
initial-state neutron-proton pairs~\cite{Piasetzky:2006ai}.

The Jefferson Lab experiment made use of the two Hall~A high resolution spectrometers~\cite{Alcorn:2004sb}
to detect the electron and proton from the $^{12}$C(e,e'p) reaction with Q$^2$ = 2 [GeV/c]$^2$ and
$x_B > 1$.   In the direction of the missing momentum vector, a proton detector known
as BigBite and the Hall A Neutron Detector, HAND, were placed in such a way that both recoiling protons and neutron could be detected in nearly the same solid angle acceptance.  While the experiment saw the signal of
correlated proton-proton pairs~\cite{Shneor:2007tu}; what they discovered was that an overwhelming faction
of the events   
were coming from initial-state neutron-proton pairs~\cite{Subedi:2008zz}.  The dominance of neutron-proton
paris over proton-proton pair was explained by 
several groups~\cite{Schiavilla:2006xx,Sargsian:2005ru,Wiringa:2008dn,Alvioli:2007zz}
as being due to short-range tensor correlations.
In 2011, the experiment will be repeated, 
but will go to even larger missing momentum.  The new experiment expects to finally 
get into exclusive kinematics where the short-range part of the nucleon-nucleon potential 
is dominating the reaction.

\section{Strength of Correlations in Nuclei}

With the exclusive data showing that the inclusive scaling is dominated 
by initial-state quasi-deuteron neutron-proton correlations, one can confidently use the 
inclusive scaling results to calculate the strength of the high-momentum part of
the momentum distribution for
various nuclei.
This is done by using a realistic nucleon-nucleon potential~\cite{Wiringa:1994wb} 
to calculate the deuteron's nucleon momentum distribution, $n_D(p)$,
where the function was normalized such 
that
\begin{equation}
1 = 4\pi \int_{0}^{\infty} p^2 \times n_D(p) \times dp
\end{equation}.
Since the plateaus in the inclusive scaling indicate that the functional forms of the
distribution must all have the same shape in this region, the relation for the high momentum 
part of the distribution can be written
\begin{equation}
\label{eq-momentum}
n_A(p) = n_D(p) \times a_{2N} 
\label{scaling-relation}
\end{equation}
where in this case $a_{2N}$ is simply the magnitude of the scaling plateau taken from data. 
To calculate the fraction of the initial-state above a certain momentum
for any nuclei where $a_{2N}$ has been measured, one simply calculates
\begin{equation}
4\pi \int_{p_{min}}^{\infty} p^2 \times n_A(p) \times dp.
\end{equation}
The integrations of $n_D(p)$ and $n_A(p)$ starting from various minimum 
momentums, $p_{min}$, are shown in Table~\ref{integration-table}. 

\begin{table}[tb]
\centering
\begin{tabular}{cccccc}
\hline
\tablehead{1}{c}{b}{Integration Lower Limit\\$[$MeV/c$]$} &
\tablehead{1}{c}{b}{Deuteron\\$[$\%$]$} &
\tablehead{1}{c}{b}{Helium-3\\$[$\%$]$} &
\tablehead{1}{c}{b}{Helium-4\\$[$\%$]$} &
\tablehead{1}{c}{b}{Carbon-12\\$[$\%$]$}  &  
\tablehead{1}{c}{b}{Iron-56\\$[$\%$]$}   \\ \hline
0         & 100\tablenote{Check of proper normalization of the deuteron momentum distribution.}    &  
-\tablenote{Many body systems are only calculated above 250~MeV/c where Eq.~\ref{scaling-relation} should be valid.}    &
- &
- \\ 
250       & 5.0    &  10   &  20    &  24     & 28      \\
275       & 4.0    &  8.0  &  16    &  19     & 22      \\
300       & 3.2    &  6.4  &  13    &  15     & 18      \\
370       & 2.5    &  5.0  &  10    &  12     & 15      \\ 
400       & 2.2    &  4.4  &  8.8   &  10     & 13      \\
440       & 1.8    &  3.6  &  7.2   &  8.6    & 10      \\
500       & 1.3    &  2.6  &  5.2   &  6.2    & 7.4      \\ \hline
\end{tabular}
\caption{The percent initial-state momentums above different minimum initial momentum values..
The Argonne v-18 nucleon-nucleon potential was used to generate the momentum distribution 
for the deuteron and the function properly normalized.
For the other nuclei, the calculation was only done with momentums greater then 250~MeV/c 
where the relation $n(p)_A = n(p)_d \times a_{2N}$ can be used and where $a_{2N}$ is 
taken from experimental $x_B > 1$ inclusive nuclear scaling ratios
of $^3$He, $^4$He, $^{12}$C, and $^{56}$Fe to deuterium of 2, 4, 4.8 and 5.7, 
respectively~\cite{Frankfurt:1993sp,Egiyan:2003vg}.}
\label{integration-table}
\end{table}

\section{The Nuclear EMC Effect}

From the results in Table~\ref{integration-table}, it is clear the magnitude of the high-momentum
tail is non-negligible fraction of the initial-state.
If this is truly an initial-state effect, then it must appear globally in nuclear physics data.
One place to look to see if these numbers provide a solution is to the long standing puzzle of
the EMC effect.

Data from the EMC collaboration showed that if one takes ratios of deep-inelastic scattering data 
from various nuclei one finds structure in the ratios~\cite{Aubert:1983xm}.  
This is very different from a simple minded
picture where one would expect the partons in nucleons to all act the same and the per nucleon ratio
to simply be one.  Many possible explanations have been put forward 
to explain this phenomenon~\cite{Geesaman:1995yd}, yet to this day there is no generally accepted solution.

\begin{figure}[htb]
  \includegraphics[width=0.45\textwidth]{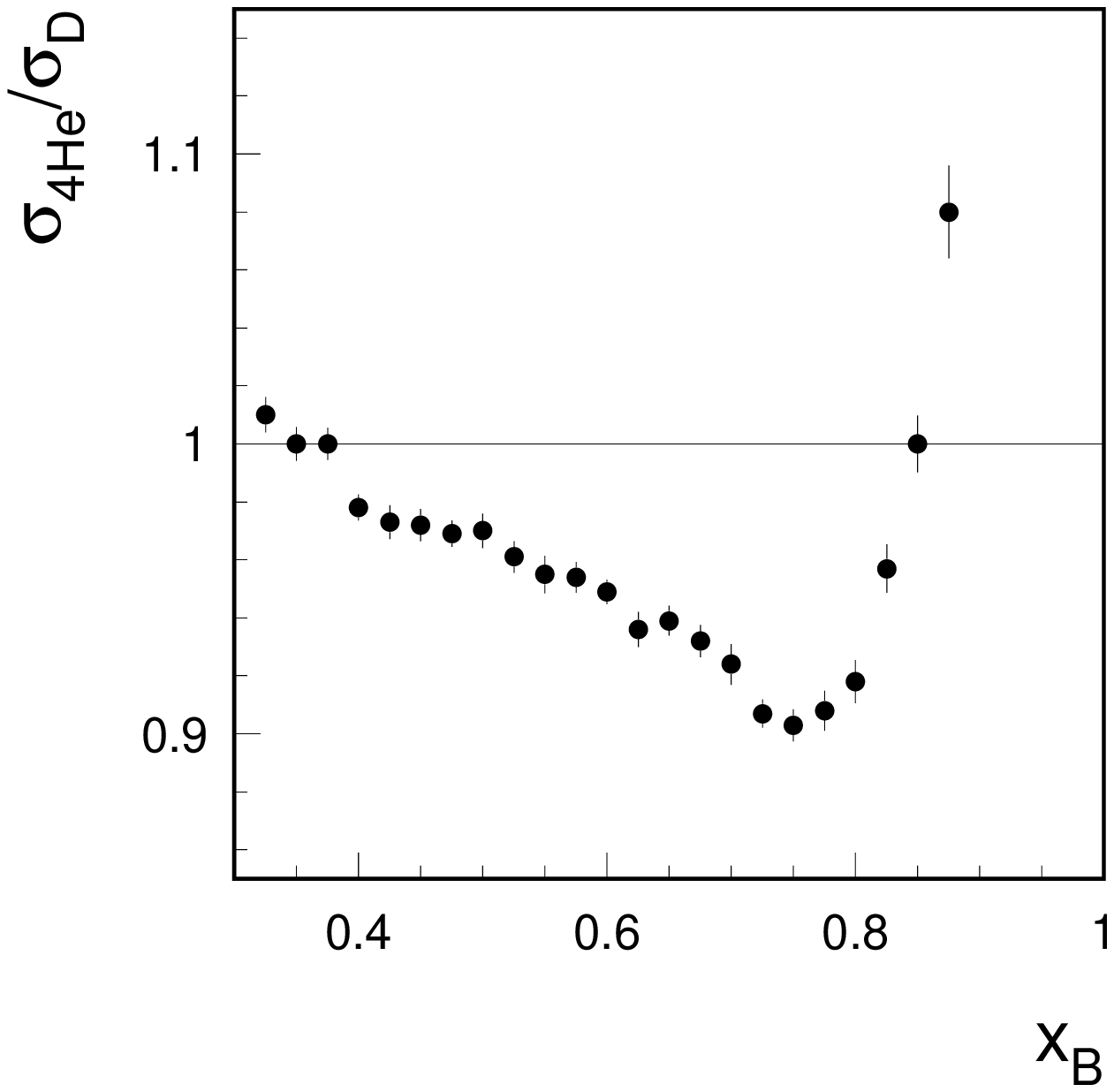}
  \hspace{5mm}
  \includegraphics[width=0.45\textwidth]{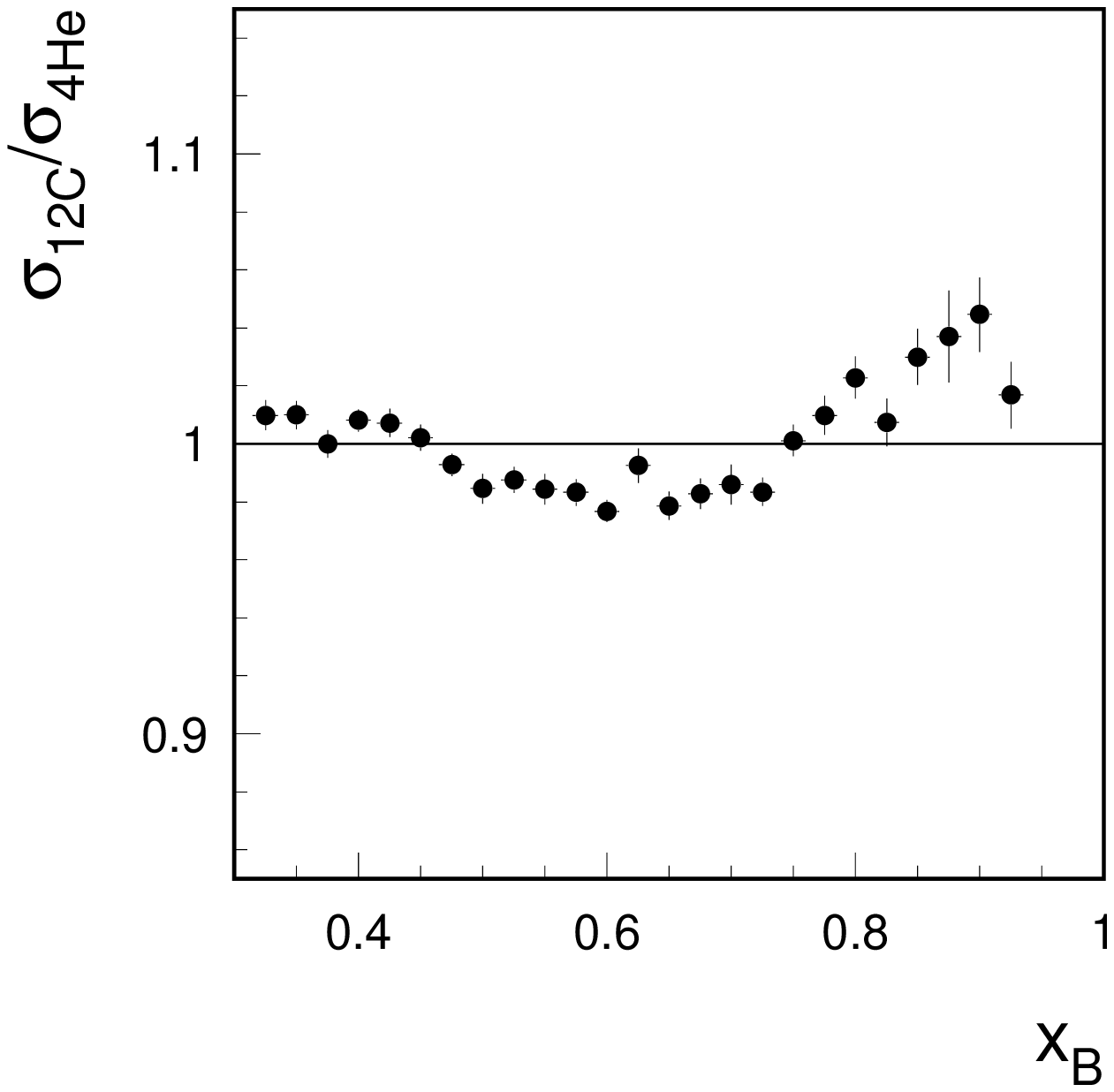}
\caption{Shown on the left is the Hall C deep inelastic per nucleon
cross section ratio of $^4$He to D with its characteristic
dip around $x_B$ of 0.75~\cite{Seely:2009gt}.   On the other hand,
the ratio of $^{12}$C to $^{4}$He the effect is relatively small, which
is consistent with the claim EMC is a local density effect.}
\label{emcs}
\end{figure}

\begin{figure}[htb]
  \includegraphics[width=\textwidth]{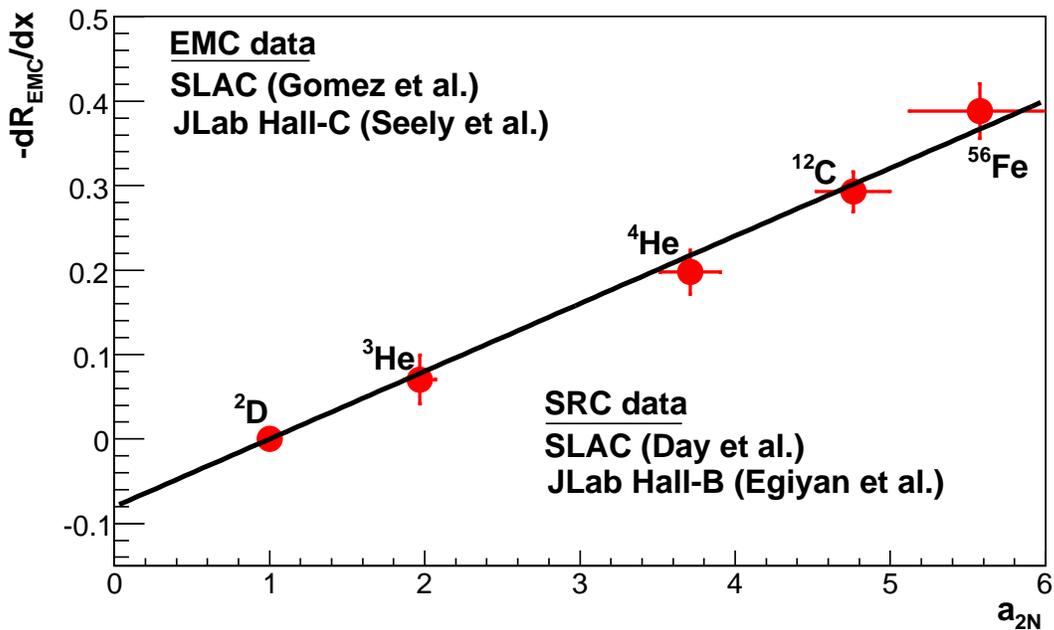}
  \caption{Shown is the $0.3 < x_B < 0.7$ EMC slopes, dR/dx, for various nuclei 
from Seely and Gomez~\cite{Seely:2009gt,Gomez:1993ri}
plotted vs the magnitude of the $x_B > 1$ nuclear scaling ratio, $a_{2N}$, for
various nuclei from Day and Egiyan~\cite{Day:1990mf,Egiyan:2003vg}.  
The linear correction between these two seemly disconnected regions
is already striking and soon to be published data will further enhance the results~\cite{Fomin:2008fq}.}
  \label{emc-src}
\end{figure}

It is interesting, is to plot EMC type data with a different nuclei in both the numerator and denominator
as shown in Fig.~\ref{emcs}.  There even though the ratio of the number of nuclei is bigger 
for $^{12}$C/$^4$He and $^{4}$He/D, 3:1 vs 2:1, the EMC is dramatically bigger for the $^4$He/D ratio.
This makes one immediately ask what is nearly the same between $^{12}$C and $^4$He, but very different
for $^4$He and D.  One obvious answer is the average binding energy; but another answer, related to the
first, is the strength of short-range correlations.
In fact, as pointed out in Refs.~\cite{Higinbotham:2010ye,Weinstein:2010rt}, 
the magnitude of the EMC effect seems to follow the magnitudes of the nuclear scaling plateaus.  
This phenomenological correlation between the EMC effect and the scaling plateaus is shown in Fig.~\ref{emc-src}.
Assuming this correlation is showing the reason for the EMC effect,
it is even possible to use the magnitudes of the scaling plateaus to correct the deuteron back to
the case of a free proton and neutron allowing for a phenomenological extraction
of the F2n structure function~\cite{Weinstein:2010rt}.

\section{Summary}

After years of trying to cleanly probe short-range correlations, recent Q$^2 > 1$~[GeV/c]$^2$
and $x_B > 1$ experiments seem to have finally isolated the signal.  The beauty is not
from any one result, but from the fact that all of the new data seems to be pointing at the
same underlying mechanism.  In fact, with the the magnitude of short-range correlations in hand,
it seems to provide a phenomenological explanation for the $0.3 < x_B < 0.7$ EMC effect slope
and provide a phenomenological way of extracting the neutron's deep inelastic structure function.


\begin{theacknowledgments}
Special thanks to Eli Piatetzky for getting me involved in short-range nucleon-nucleon studies
and also for suggesting to the organizers I present this talk.
I also thank Javier Gomez, Larry Weinstein, Mark Strikman, Jerry Miller, Franz Gross, 
Wally van Orden, and Misak Sargsian for the many interesting
discussions about what these new experimental results are trying to tell us.
This work was supported by the U.S.\ Department of Energy
and Jefferson Science Associates which operates
the Thomas Jefferson National Accelerator Facility under DOE
contract DE-AC05-06OR23177.

\end{theacknowledgments}

%

\end{document}